\documentclass[aps,preprint]{revtex4}%
\usepackage{amsfonts}
\usepackage{amsmath}
\usepackage{amssymb}
\usepackage{graphicx}%
\begin{document}
\title[ ]{Asymmetric propagation of electronic wave function through molecular bonding
and anti-bonding states.}
\preprint{ }
\author{Muhammad Imran$^{\ast}$}
\affiliation{Department of Physics, Quaid-i-Azam University, Islamabad, Pakistan.r}

\begin{abstract}
Electron transport through molecular bridge shows novel quantum features.
Propogation of electronic wave function through molecular bridge is completely
different than individual atomic bridge employed between two contacts. In case
of molecular bridge electronic wave propagators interfere and effect
conduction through molecular bonding and anti-bonding states.In the present
work i showed through simple calculation that interference of electronic wave
propagators cause asymmetric propagation of electronic wave through bonding
and anti-bonding state. While for hydrogenic molecule these propagators
interfere completely destructively for bonding state and constructively for
anti-bonding state, giving rise to only one peak in spectral function for
anti-bonding state.

\end{abstract}
\volumeyear{year}
\volumenumber{number}
\issuenumber{number}
\eid{identifier}
\date[Date text]{date}
\received[Received text]{date}

\revised[Revised text]{date}

\accepted[Accepted text]{date}

\published[Published text]{date}

\startpage{1}
\endpage{2}
\maketitle

Aptitude of semiconductor technology towards miniaturization of electronic
devices has lead to emergence of nano and molecular electronics. Now the work
of nano scientists is to control electron transport at such minute scale. As
the size of the nano devices are comparable to the wavelength of current
carriers so quantum features are playing dominant role in nano devices. A
theoretical model of electron transport through a molecule was first presented
by A.Aviram and Mark.A.Ratner\cite{1,2}. Since then there has been great
interest both in experimental and theoretical sides for progress of molecular
electronics\cite{15,16,17}. A single molecule could be used as electronic
mixers\cite{10}, switches\cite{11,12}, and rectifiers\cite{13}. Along with
progress of molecular electronics a lot of problems are appearing both from
experimental and theoretical sides. For instance, from experimental side its
very difficult to place a molecule between unambiguous contacts\cite{14}.
Electron transport through molecule has very non-intuitive characteristics.
Many experimentalists and theoreticians of nano science refers to hydrogenic
molecular transport to be explained by transport through a single
channel\cite{3,4,5,6}. K. S. Thygesen et.al \cite{7} presented that
transmission through molecular anti-bonding state becomes nearly equal to one,
where as they utilize density function theory with wannier function. They
predicted that a hydrogen molecule contacted by platinum contacts will provide
only one resonant robust channel for conduction which is anti-bonding state of
molecule. Apart from hydrogenic molecule, even more complex molecules shows
asymmetry in conduction through Highest occupied molecular orbital (HOMO) and
lowest unoccupied molecular orbital (LUMO)\cite{9}. In the present work I
explain a possible physical mechanism which could cause this asymmetry in
conduction through a diatomic molecular bonding and anti-bonding states.

In modeling Hamiltonian for the diatomic molecule employed between contacts I
considered contacts as charge carriers bath and electron on diatomic molecule
loss energy with it. Therefore I take complex energy for electron on each atom.%

\begin{equation}
H=\sum_{i=1}^{2}\tilde{\epsilon}_{i}d_{i}^{\dag}d_{i}+\sum_{i,j=1i\neq j}%
^{2}\tau d_{i}^{\dag}d_{j} \label{1}%
\end{equation}

The present Hamiltonian is written under approximation of linear combination
of atomic orbitals and in second quantized form\cite{8}. Here $d_{i}^{\dag
}(d_{i})$ represents creation(annihilation) of electron on ith atom, where as
second term refers to inter-atomic coupling and $\tau$ shows inter-atomic
coupling energy. Where as \textquotedblleft\ $\tilde{\epsilon}_{i}%
=\epsilon_{i}-i\Gamma$ \textquotedblright\ $\epsilon_{i}$ gives electronic
energy of the ith atom electron and $\Gamma$ shows dissipation of electron
energy with contacts.

Now time evolution of \ operator is given by Heisenberg's equation of motion,
(I assume for simplicity $\hbar=1$ )%

\begin{equation}
i\dfrac{\partial}{\partial t}[Exp(i\tilde{\epsilon}_{_{1}}t)d_{1}(t)]=\tau
Exp(i\tilde{\epsilon}_{_{1}}t)d_{2}(t) \label{2}%
\end{equation}

\begin{equation}
i\dfrac{\partial}{\partial t}[Exp(i\tilde{\epsilon}_{_{2}}t)d_{2}(t)]=\tau
Exp(i\tilde{\epsilon}_{2}t)d_{1}(t) \label{3}%
\end{equation}

\begin{equation}
F_{1}(t)\equiv Exp(i\tilde{\epsilon}_{_{1}}t)d_{1}(t) \label{4}%
\end{equation}

\begin{equation}
F_{2}(t)\equiv Exp(i\tilde{\epsilon}_{2}t)d_{2}(t) \label{5}%
\end{equation}

\begin{equation}
\frac{i}{\tau}Exp(i\Delta t)\dfrac{\partial}{\partial t}F_{1}(t)=F_{2}(t)
\label{6}%
\end{equation}

\begin{equation}
\frac{i}{\tau}Exp(-i\Delta t)\dfrac{\partial}{\partial t}F_{2}(t)=F_{1}(t)
\label{7}%
\end{equation}
Here \textquotedblleft$\Delta=(\epsilon_{_{2}}-\epsilon_{_{1}})$
\textquotedblright. For decoupling equations (6 \&7) I operate $i\dfrac
{\partial}{\partial t}$ on above two equations,%

\begin{equation}
\ddot{F}_{1}+i\Delta\dot{F}_{1}+\tau^{2}F_{1}=0 \label{8}%
\end{equation}

\begin{equation}
\ddot{F}_{2}-i\Delta\dot{F}_{2}+\tau^{2}F_{2}=0 \label{9}%
\end{equation}
Here $\bullet\&$ $(\bullet\bullet)$ represents $\dfrac{\partial}{\partial
t}\&$ $(\dfrac{\partial^{2}}{\partial t^{2}})$ ,

By solving above two differential equations,%

\begin{align}
d_{1}(t)  &  =\frac{1}{2\gamma}[\{(\gamma+\Delta)d_{1}(0)-2\tau d_{2}%
(0)\}Exp(-i(\epsilon_{b}-i\Gamma)t)+\label{10}\\
&  \{(\gamma-\Delta)d_{1}(0)+2\tau d_{2}(0)\}Exp(-i(\epsilon_{a}%
-i\Gamma)t)\}]\nonumber
\end{align}

\begin{align}
d_{2}(t)  &  =\frac{1}{2\gamma}[\{(\gamma-\Delta)d_{2}(0)-2\tau d_{1}%
(0)\}Exp(-i(\epsilon_{b}-i\Gamma)t)+\label{11}\\
&  \{(\gamma+\Delta)d_{2}(0)+2\tau d_{1}(0)\}Exp(-i(\epsilon_{a}%
-i\Gamma)t)\}]\nonumber
\end{align}
Here \textquotedblleft$\gamma=\sqrt{\Delta^{2}+4\tau^{2}}$ , $\epsilon
_{b}=\dfrac{(\epsilon_{1}+\epsilon_{2}-\gamma)}{2}$ \ and $\epsilon_{a}%
=\dfrac{(\epsilon_{1}+\epsilon_{2}+\gamma)}{2}$ \ \textquotedblright, Where as
$\epsilon_{b}(\epsilon_{a})$ represents molecular bonding (molecular
anti-bonding) energy.

Similarly,%

\begin{align}
d_{1}^{\dag}(t)  &  =\frac{1}{2\gamma}[\{(\gamma+\Delta)d_{1}^{\dag}(0)-2\tau
d_{2}^{\dag}(0)\}Exp(i(\epsilon_{b}-i\Gamma)t)+\label{12}\\
&  \{(\gamma-\Delta)d_{1}^{\dag}(0)+2\tau d_{2}^{\dag}(0)\}Exp(i(\epsilon
_{a}-i\Gamma)t)\}]\nonumber
\end{align}

\begin{align}
d_{2}^{\dag}(t)  &  =\frac{1}{2\gamma}[\{(\gamma-\Delta)d_{2}^{\dag}(0)-2\tau
d_{1}^{\dag}(0)\}Exp(i(\epsilon_{b}-i\Gamma)t)+\label{13}\\
&  \{(\gamma+\Delta)d_{2}^{\dag}(0)+2\tau d_{1}^{\dag}(0)\}Exp(i(\epsilon
_{a}-i\Gamma)t)\}]\nonumber
\end{align}
Finally, propagator for diatomic molecular system will be,%

\begin{equation}
G^{R}\left(  t,t^{\prime}\right)  =\sum_{i,j=1}^{2}G_{ij}^{R}\left(
t,t^{\prime}\right)  \label{14}%
\end{equation}
Here \textquotedblleft\ $G^{R}\left(  t,t^{\prime}\right)  $
\textquotedblright\ represents retarded Green's function,%

\begin{equation}
G_{ij}^{R}\left(  t,t^{\prime}\right)  \equiv-i\theta(t-t^{\prime
})\left\langle [d_{i}(t)\text{ \ }d_{j}^{\dagger}\left(  t^{\prime}\right)
]\right\rangle \label{15}%
\end{equation}
Where as, in frequency space,%

\begin{equation}
G_{ij}^{R}\left(  \epsilon\right)  =\int d(t-t^{\prime})Exp(i\epsilon
(t-t^{\prime}))G_{ij}^{R}\left(  t,t^{\prime}\right)  \label{16}%
\end{equation}
By utilizing eqs (10-13),%

\begin{equation}
G_{11}^{R}\left(  \epsilon\right)  =\frac{1}{2\gamma}[\dfrac{(\gamma+\Delta
)}{(\epsilon-\epsilon_{b}+i\Gamma)}+\dfrac{(\gamma-\Delta)}{(\epsilon
-\epsilon_{a}+i\Gamma)}] \label{17}%
\end{equation}

\begin{equation}
G_{12}^{R}\left(  \epsilon\right)  =\dfrac{\tau}{\gamma}[\dfrac{1}%
{(\epsilon-\epsilon_{a}+i\Gamma)}-\dfrac{1}{(\epsilon-\epsilon_{b}+i\Gamma)}]
\label{18}%
\end{equation}

\begin{equation}
G_{21}^{R}\left(  \epsilon\right)  =\dfrac{\tau}{\gamma}[\dfrac{1}%
{(\epsilon-\epsilon_{a}+i\Gamma)}-\dfrac{1}{(\epsilon-\epsilon_{b}+i\Gamma)}]
\label{19}%
\end{equation}

\begin{equation}
G_{22}^{R}\left(  \epsilon\right)  =\frac{1}{2\gamma}[\dfrac{(\gamma-\Delta
)}{(\epsilon-\epsilon_{b}+i\Gamma)}+\dfrac{(\gamma+\Delta)}{(\epsilon
-\epsilon_{a}+i\Gamma)}] \label{20}%
\end{equation}
Now the spectral function for the system will be,%

\begin{equation}
A\left(  \epsilon\right)  =-2\sum_{ij}\operatorname{Im}[G_{ij}^{R}\left(
\epsilon\right)  ] \label{21}%
\end{equation}
The difference in bonding and anti-bonding energies for diatomic molecule
(with both atoms of same energy) increase linearly with increase in
inter-atomic coupling\textquotedblleft\ $\epsilon_{b}=\epsilon_{0}-\tau$ $\&$
$\epsilon_{a}=\epsilon_{0}+\tau$ \textquotedblright. While for diatomic
molecule( with both atoms of different energy) the difference in bonding and
anti-bonding energy increases in terms of square roots \textquotedblleft%
\ $\epsilon_{b}=\dfrac{(\epsilon+\epsilon_{1}-\sqrt{\Delta^{2}+4\tau^{2}})}%
{2}$ $\&$ $\epsilon_{a}=\dfrac{(\epsilon+\epsilon_{2}+\sqrt{\Delta^{2}%
+4\tau^{2}})}{2}$ \textquotedblright\ \ In Fig(1) it is shown that difference
in energy with increase in inter-atomic coupling for bonding and anti-bonding
energy changes differently for two atoms of same energy level and two atoms of
different energy level.

%

\begin{figure}
[ptb]
\begin{center}
\includegraphics[
height=3.3287in,
width=5.3195in
]%
{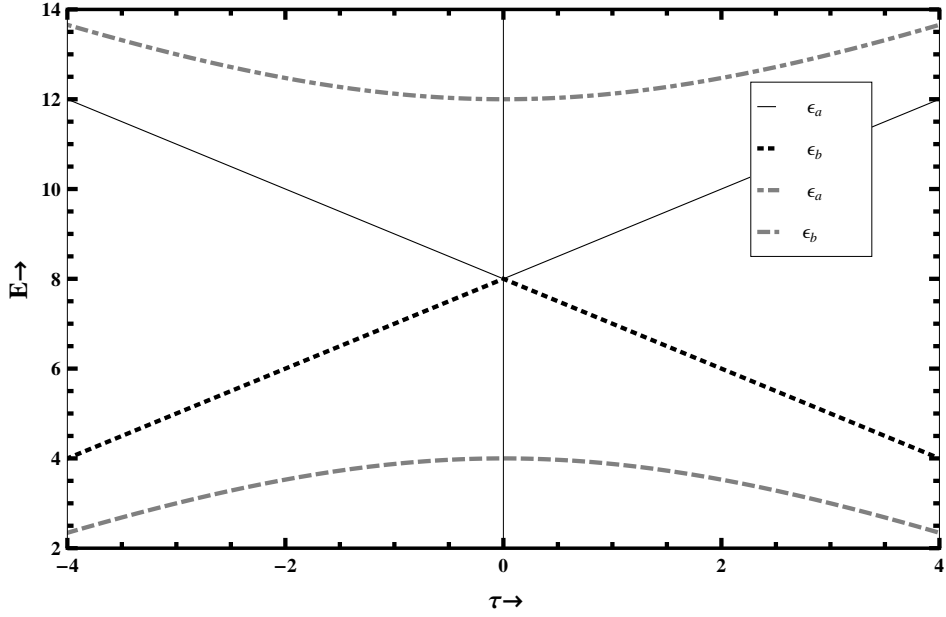}%
\caption{Energy difference between bonding and anti-bonding energies increase
linearly for two atoms of same energy level $\epsilon_{1}=\epsilon_{2}=8$
,While it increases in terms of square root for two atoms of different energy
level $\epsilon_{1}=4,$ $\epsilon_{2}=12$ .}%
\end{center}
\end{figure}

In Fig(2) spectral function solid black peak refers to two atoms of same
energy levels with no inter-atomic coupling \textquotedblleft\ $A(\epsilon
)=\dfrac{4\Gamma}{(\epsilon-\epsilon_{0})^{2}+\Gamma^{2}}$ \textquotedblright,
which gets peak value at \textquotedblleft$\epsilon=\epsilon_{0}$
\textquotedblright\ . With finite value of inter-atomic coupling only single
peak appears in spectral function which is shifted to right by amount
\textquotedblleft$\epsilon_{0}+\tau$\textquotedblright\ .This peak correspond
to anti-bonding state \textquotedblleft$A(\epsilon)=\dfrac{4\Gamma}%
{(\epsilon-\epsilon_{a})^{2}+\Gamma^{2}}$ \textquotedblright\ while no peak of
bonding state appears as propagators destructively interfere to cancel bonding
state peak. With increase in inter-atomic coupling the anti-bonding state peak
shifts right by amount \textquotedblleft\ $\tau$\textquotedblright\ . This
provides a good physical explanation for hydrogenic molecular conduction
through\ a single channel of anti-bonding state, as predicted by K. S.
Thygesen et.al \cite{7}, although I have not included contacts and molecular
coupling geometry in our model.%

\begin{figure}
[ptb]
\begin{center}
\includegraphics[
height=3.3287in,
width=5.3195in
]%
{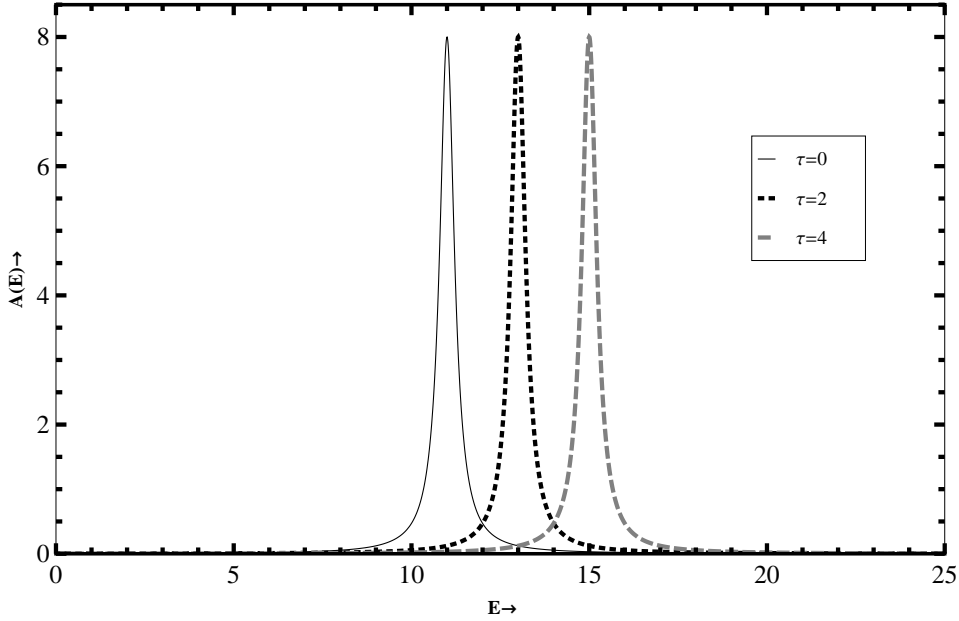}%
\caption{Spectral function for a diatomic molecule consisting of two atoms of
the same energy level $\epsilon_{1}=\epsilon_{2}=11$ and dissipation of
electron energy with contacts $\Gamma=0.25$ .}%
\end{center}
\end{figure}

A diatomic molecule is completely different than two atoms laterally placed
between contacts without any inter-atomic coupling. As in former case
electronic wave function get hybridized and collectively two atoms propagator
amplitudes interfere. In spectral function two symmetric peaks will be seen in
case of two atoms with no inter atomic coupling \textquotedblleft\ $A\left(
\epsilon\right)  =\dfrac{2\Gamma}{(\epsilon-\epsilon_{1})^{2}+\Gamma^{2}%
}+\dfrac{2\Gamma}{(\epsilon-\epsilon_{2})^{2}+\Gamma^{2}}$ \textquotedblright%
\ which gets peak value at \textquotedblleft$\epsilon=\epsilon_{1}$ $\&$
$\epsilon=\epsilon_{2}$ \textquotedblright\ \ while in a diatomic molecule
rather than two symmetric peaks we get two asymmetric peaks for bonding and
anti-bonding states and its amplitude is also different than two atoms with no
inter-atomic coupling peaks.

In Fig(3) spectral function for a diatomic molecule consisting of two atoms of
different energy level is shown. Here two symmetric peaks in spectral function
refers to no inter-atomic coupling, while with finite value of inter-atomic
coupling two peaks start shifting with asymmetric amplitude. Here bonding
state peak amplitude is smaller than two atoms with no inter atomic coupling
because of destructive interference between propagators, while anti bonding
state peak amplitude is larger than two atoms with no inter atomic coupling
peaks in spectral function by the amount the propagators interfere
constructively. With increase in inter-atomic coupling separation between
bonding and anti-bonding states peaks start increasing and asymmetry in their
peaks amplitude also increase.%

\begin{figure}
[ptb]
\begin{center}
\includegraphics[
height=3.3287in,
width=5.3195in
]%
{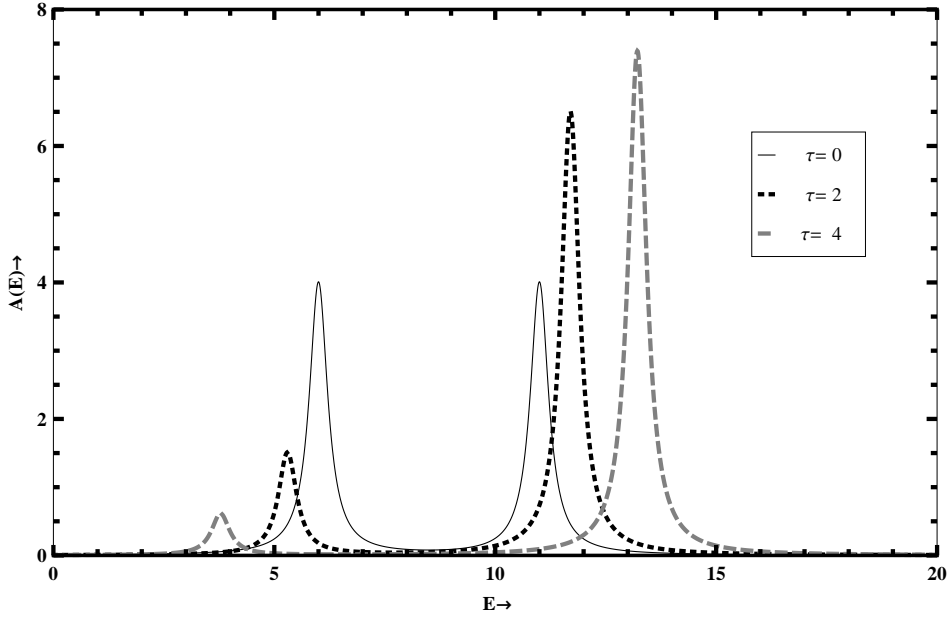}%
\caption{Spectral function for diatomic molecule consisting of two atoms of
different energy level $\epsilon_{1}=6$ , $\epsilon_{2}=11$ and dissipation of
electron energy with contacts $\Gamma=0.25$.}%
\end{center}
\end{figure}

Molecular bonding state is symmetric state and it tends to localize charge
carriers while molecular anti-bonding state is anti-symmetric state it tends
to delocalize charge carriers. This effects electronic wave propagators in a
way that electronic wave propagation amplitude interferes constructively for
anti-bonding state which gives large amplitude peak in spectral function while
electronic wave propagation amplitudes interfere destructively for bonding
state and gives smaller peak amplitude in spectral function for bonding state.

$^{\ast}$imran1gee@gmail.com

\end{document}